\documentclass[aps,prx,twocolumn,groupedaddress,showpacs,floatfix,superscriptaddress,longbibliography]{revtex4-2}
\usepackage[plainpages=false,pdfpagelabels,colorlinks=true,linkcolor=red,urlcolor=blue,citecolor=blue,pdftitle={Title},pdfauthor={},pdfdisplaydoctitle=true,pdfduplex=DuplexFlipLongEdge]{hyperref}
\usepackage{epsfig}
\usepackage{amsmath,amssymb}
\usepackage{physics}
\usepackage{bm}
\usepackage{graphicx}
\usepackage[dvipsnames,usenames]{color}
\usepackage[normalem]{ulem}
\usepackage{soul} % to use \st for strikeout

\newcommand{\tcb}{\textcolor{blue}}
\definecolor{darkred}{rgb}{0.80,0,0}
\definecolor{blood}{rgb}{0.50,0,0}
\definecolor{brightred}{rgb}{1,0,0}
\definecolor{orange}{rgb}{1,0.3,0}
\definecolor{bluegreen}{rgb}{0,0.5,0.5}
\definecolor{lightblue}{rgb}{0,0.5,0.8}
\definecolor{darkgreen}{rgb}{0,0.5,0}
\definecolor{green}{rgb}{0,0.70,0}
\definecolor{darkblue}{rgb}{0,0,0.80}
\definecolor{magenta}{rgb}{1,0,1}
\definecolor{softmagenta}{rgb}{0.85,0.1,0.6}
\definecolor{mauve}{rgb}{0.6,0.1,1}
\definecolor{white}{rgb}{1,1,1}
\definecolor{black}{rgb}{0,0,0}
\graphicspath{{Images/}}

\begin{document}

\title{Effect of Emitters on Quantum State Transfer in Coupled Cavity Arrays}
\author{Eli Baum} 
\affiliation{Department of Physics, University of California, Davis, CA 95616,USA}
\author{Amelia Broman} 
\affiliation{Department of Physics and Astronomy, Carleton College,
Olin Hall, 215 Goodsell Circle,
Northfield, MN 55057 USA}
\author{Trevor Clarke} 
\affiliation{Department of Physics, University of California, Davis, CA 95616,USA}
\author{Natanael C. Costa}
\affiliation{Instituto de F\'isica, Universidade Federal do Rio de Janeiro, Cx.P. 68.528, 21941-972, Rio de Janeiro RJ, Brazil}
\author{Jack Mucciaccio} 
\affiliation{Department of Physics, Coe College, Cedar Rapids, IA 52402, USA}
\author{Alexander Yue} 
\affiliation{Department of Physics, University of California, Davis, CA 95616,USA}
\author{Yuxi Zhang}
\affiliation{Department of Physics, University of California, Davis, CA 95616,USA}
\author{Victoria Norman}
\affiliation{Department of Physics, University of California, Davis, CA 95616,USA}
\affiliation{Department of Electrical and Computer Engineering, University of California, Davis, CA 95616,USA}
\author{Jesse Patton}
\affiliation{Department of Electrical and Computer Engineering, University of California, Davis, CA 95616,USA}
\author{Marina Radulaski}
\affiliation{Department of Electrical and Computer Engineering, University of California, Davis, CA 95616,USA}
\author{Richard T. Scalettar}
\affiliation{Department of Physics, University of California, Davis, CA 95616,USA}

\date{\today}

\begin{abstract}
Over the last decade, conditions for
 perfect state transfer in quantum spin chains
have been discovered, and 
 their experimental realizations addressed.
 In this paper, we consider an
 extension of such studies to quantum state transfer in
 a coupled cavity array including the effects of atoms 
in the cavities which can absorb and emit photons as they
 propagate down the array. Our model is equivalent to
 previously examined spin chains in the one-excitation
 sector and in the absence of emitters. We introduce a
 Monte Carlo approach to the inverse eigenvalue problem
 which allows the determination of the inter-cavity
 and cavity-emitter couplings resulting in near-perfect
 quantum state transfer fidelity, and examine the
 time dependent polariton wave function through exact
diagonalization of the resulting Tavis-Cummings-Hubbard Hamiltonian.
The effect of inhomogeneous emitter locations is also evaluated.
\end{abstract}

\maketitle

%%%%%%%%%%%%%%%%%%%%%%%%%%%%%%%%%%%%%%%%
%%%%%%%%%%%%%%%%%%%%%%%%%%%%%%%%%%%%%%%%
\section{Introduction}\label{sec:Intro}
%%%%%%%%%%%%%%%%%%%%%%%%%%%%%%%%%%%%%%%%
%%%%%%%%%%%%%%%%%%%%%%%%%%%%%%%%%%%%%%%%

\par One of the most important questions in quantum information theory is the 
faithful and rapid, transmission of a quantum state from one location
to another. Quantum spin chains have proven to be a very useful
and powerful context in which to explore fundamental issues, including
the possibility of perfect transfer, the effect of disorder,
and the interplay between high fidelity and speed of propagation
 \cite{bose03,christandl2004,christandl2005,karbach2005,bose07,felicetti2014,almeida2016}.
In the case where a single excitation is present (one up spin
in a background of down spins) the resulting Hamiltonian is represented
by a tridiagonal (`Jacobi') matrix.
A general classification of the eigenspectra of such matrices
which result in perfect Quantum  State Transfer (QST) has emerged,
as has the determination of the requisite 
`fully engineered' intersite exchange
constants $J_{i}$ \cite{christandl2004,christandl2005}.  Interestingly, it was also found that nearly 
perfect QST could be achieved with more limited and feasible
`boundary' engineering in which the $J_{i}$ are uniform in the interior,
and take special values only at the beginning and end of the chain  \cite{banchi2013}.
Although boundary engineering has the advantage of requiring less 
precise, and therefore less experimentally challenging, tuning,
good QST is achieved only in the limit of weak coupling at the ends,
and hence is compromised by long transfer times.

\par A subsequent focus was on the effect of disorder on QST, since in any 
physical realization a certain degree of randomness is inevitable.  
There are many eigenvalue distributions which give rise to
perfect QST in the ideal limit, and therefore
one line of investigation concerned the types of such engineered
spectra which are most robust to disorder \cite{zwick11}. 
A key observation was that once randomness is present, the resulting
degradations of state transfer of fully and boundary 
engineered chains are roughly
similar, so that there is limited incentive to attempt full engineering
as far as fidelity itself is concerned \cite{zwick15}.
(The problem of longer transfer times in boundary engineered chains, however,
remains.)

\par
\begin{figure}[t] 
\includegraphics[width=2.0in,height=1.0in,angle=0]{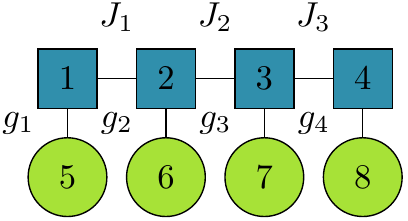}
\caption{ Geometry of the one dimensional Jaynes-Cummings-Hubbard Hamiltonian
with $N=4$ cavities each containing $M_i=1$ emitter. Photons can `hop' between
 a set of cavities $i$ and $i+1$ (squares) via the inter-cavity
 coupling rates
 $J_{i}$. Within each cavity $i$ a photon can be absorbed and excite the emitter
 (circles) via $g_{i}$. The same process can be reversed, traveling from emitter
to cavity also via $g_{i}$.  The numbers indicate our (arbitrary) convention
 for labeling the states in the
Hilbert space, i.e. the rows of our matrix in the single excitation sector.}
\label{fig:cavemt}
\end{figure}

 In this paper, we consider QST
within a different physical and geometric context, namely when
the `backbone' chain also possesses branches to localized qubits,
forming a `comb-like' geometry as illustrated in 
Fig.~\ref{fig:cavemt}.
We are motivated by the study of the nature and propagation of
excitations in a coupled cavity array (CCA)  \cite{hartmann08,angelakis07,tomadin10}.
A CCA consists of a chain of optical
cavities, which might be empty or may contain one or more atom-like
emitters coupled to the cavity’s electromagnetic field. 
Photons hop between adjacent cavities in the CCA due to the
overlap of neighboring resonance modes,
and strong interactions between light and matter can be induced.
These emitters form the `rungs' which dress our one-dimensional chain of cavities.

\par CCAs have become increasingly experimentally viable in
recent years \cite{hartmann16,majumdar12}, and have been especially
intriguing as possible venues for exploring superfluid to
insulator transitions and other many-body phenomena. 
However, in order to observe  such effects,
the CCA must exist in the strong coupling 
regime of cavity quantum electrodynamics, where light-matter 
interactions are stronger than losses to the environment. 
Modern integrated optical cavities achieve this by localizing light
on the (sub)wavelength %nanometer
scale. One of the commonly used optical 
resonators for these studies is the photonic crystal cavity, 
formed by periodic refractive index alteration at the %sub-wavelength 
nanoscale. 

\par One of the attractive choices for quasi-atoms which might interact
with such solid-state CCAs are color
centers formed as lattice defects in semiconductors \cite{radulaski17}. 
The defect causes electron wavefunctions
to localize at that point, effectively creating an isolated two-level system 
within a solid-state material. The most common
material substrates for this purpose are silicon carbide \cite{koehl2011room,widmann2015coherent,lukin20204h, radulaski2017scalable,babin21,majety21} and diamond \cite{hepp2014electronic, sipahigil2014indistinguishable, zhang2018strongly, bradac2019quantum}. 
%For example, \cite{majety21} describes a photonic crystal cavity with a triangular cross-section, designed for a silicon carbide substrate with various possible color centers.

\par An immediate question is whether perfect QST is still possible in these more
complex `two component' systems, and,
if so, what are the associated cavity-cavity and cavity-emitter couplings.
A second question pertains to the effect of a fundamentally different
type of `geometric' disorder which arises from inhomogeneity in the emitter
numbers and locations, rather than the previously
explored situations where randomness is introduced via bond-dependent
couplings in a fixed and regular geometry. 
A cavity which is absent an emitter corresponds,
for example,
to a missing `tooth' at that location of the comb.  We will describe the
consequences of such disorder on QST.

\par One interesting aspect of such cavity-emitter arrays is as a novel
realization of `boundary engineering'.  If atoms are placed only in the 
initial and final positions of $N-2$ cavities, the geometry is identical
as that of an $N$ cavity chain.  The emitter-cavity couplings 
then play the role of the bond strengths $J_{1}$
and $J_{N-1}$ of a spin chain.

\par A final avenue of investigation described here concerns the case of 
multiple excitations in cavity-emitter systems.  In the absence of
emitters, the Hamiltonian is quadratic and describes a set 
of independent bosonic particles (photons).  As a consequence, perfect
QST in the single excitation sector guarantees the same occurs 
for multiple excitations.  When emitters are present, the Hamiltonian
remains quadratic in the photon and emitter operators.  However
the mixed nature of the commutation relations/allowed `occupations'
makes the multi-excitation sector fundamentally different
from single excitations.  We will describe the prospects for achieving
high fidelities in this situation.

\par Our paper is organized as follows. 
In Section \ref{sec:Model} we review the
Jaynes-Cummings-Hubbard Hamiltonian (JCHH) and its matrix representation 
in the single excitation sector.   We also briefly describe the exact
diagonalization method used in the time evolution of states
and the Monte Carlo approach used to solve the inverse eigenvalue problem.
Section \ref{sec:QST-JCHH} presents evidence for the possibility of perfect QST in
cavity-emitter systems.  
These results provide `full engineering' solutions to perfect QST in the JCHH,
generalizing known spin chain results.  
Having established perfect QST in this more complex setting, we next consider,
in Sec.~\ref{sec:QST-JCHH-dis}, the effects of disorder.
Section \ref{sec:JCHHBE} and Sec.~\ref{sec:Multi}
discuss how cavity emitter systems can provide a novel realization of
boundary engineering, and the nature of QST when multiple excitations are present,
respectively.
A brief overview of experimental parameters in CCA in silicon carbide
with
color centers serving as emitters is contained in Sec.~\ref{sec:exptparam}.
Finally our results are summarized in Sec.~\ref{sec:Conclusions}.
Several details are discussed in the Appendix.

%%%%%%%%%%%%%%%%%%%%%%%%%%%%%%%%%%%%%
%%%%%%%%%%%%%%%%%%%%%%%%%%%%%%%%%%%%%
\section{Model and Methodology}\label{sec:Model}
%%%%%%%%%%%%%%%%%%%%%%%%%%%%%%%%%%%%%
%%%%%%%%%%%%%%%%%%%%%%%%%%%%%%%%%%%%%

\par The cavity-emitter arrays we will study are described by the Jaynes-Cummings-Hubbard Hamiltonian,
\begin{eqnarray}
{\cal H} &= \sum\limits_{i=1}^{N}  \Omega_i^{\phantom{\dagger}} a_{i}^{\dagger} a_{i}^{\phantom{\dagger}}
+ \sum\limits_{i=1}^{N-1}  J_{i}^{\phantom{\dagger}} \big( \, a_{i+1}^{\dagger} a_{i}^{\phantom{\dagger}}
+  a_{i}^{\dagger} a_{i+1}^{\phantom{\dagger}} \, \big)
\nonumber \\
&+\sum\limits_{i=1}^{N} \, \sum\limits_{j=1}^{M_i} \omega_{ij}^{\phantom{\dagger}} \sigma_{ij}^{+} \sigma_{ij}^{-} +
g_{ij} \big( \, a_{i}^{\dagger}  \sigma_{ij}^{-} 
+ \sigma_{ij}^{+} a_{i}^{\phantom{\dagger}} \, \big) 
\label{eq:JCHH}
\end{eqnarray}
Here $N$ is the number of cavities and $\{M_i\}$ are the numbers of emitters in cavity $i$.
$a_{i}^{\dagger}\big(a_{i}^{\phantom{\dagger}} \big)$
are photon creation (annihilation) operators in cavity $i$, and 
$\sigma_{ij}^{+}\big(\sigma_{ij}^{-} \big)$
are excitation (de-excitation) operators for emitter $j$ in cavity $i$.
The model is parameterized by cavity energies $\Omega_i$,
photon hopping rates
$J_{i}$, emitter energy levels $\omega_{ij}$ and photon-emitter 
coupling rates $g_{ij}$.
We focus on the case when there is at most one emitter per cavity,
$M_i = 0,1$, and hence will simplify the notation to $g_i$ and $\omega_i$, dropping the $j$
subscript which distinguishes different emitters in the same cavity.
In cases when the number of emitters varies, we will refer to the 
{\it sparse} JCHH.

Real cavities and emitters have finite linewidth, representing the possibility
of loss.  High quality (small linewidth) cavities and emitters are
increasingly available \cite{radulaski17}.  
Hence these effects are ignored in the present work.

\par A basis for the Hilbert space in the single excitation sector and in the absence of
emitters is the collection of states
$| \, 0 \, 0 \, 0 \, \cdots 0 \, 1_i \, 0 \cdots 0 \, \rangle$
with a single phonon in cavity $i$.  The Hamiltonian is represented by
 the tridiagonal (`Jacobi') matrix,
\begin{equation} \label{eq:hamiltonian_cav}
{{\cal H} = \begin{pmatrix}
\Omega_1 & -J_{1} & 0 & \ldots & 0\\
-J_{1} & \Omega_2 & -J_{2} & \ldots & 0\\
0 & -J_{2} & \Omega_3  & \ldots & 0\\
\vdots & \vdots & \vdots & \ddots & -J_{N-1}\\
0 & 0 & 0 & -J_{N-1} & \Omega_N  \,\, .
\end{pmatrix}}
\end{equation}

\par We compute the time evolution from an initial state $|\, \Psi(t=0) \, \rangle$
by diagonalizing ${\cal H}=S\, D S^{\dagger}$, exponentiating 
${\cal H}$ to obtain
 $U=e^{-i {\cal H} t}=S\, e^{- i D t } S^{\dagger}$, thereby finding
\begin{equation} \label{eq:time_evolution}
\ket{\psi(t)} = e^{-i {\cal H} t} \ket{\psi(0)}
\end{equation}
where we take $ \hbar = 1$. 
  We begin our system with 
  $\ket{\psi(0)} = \ket{1, 0, 0, \ldots, 0}$ 
 corresponding to a single photon contained entirely in
  cavity $i=1$ at time $t=0$ and let the system evolve in time. 
  We are interested in a final state 
  $\ket{\psi_{\rm f} } = \ket{0, 0, 0, \ldots, 1}$ 
  with the photon in cavity $i=N$.
  
\par We define the \textit{fidelity} ${\cal F}$ to be ${\cal F}
= {\rm max}_t \, f(t)$ where
$f(t) \equiv | \mel{\psi_{\rm f} }{e^{-i {\cal H} t}}{\psi(0)}|^2$ 
is the probability
 the excitation, beginning in cavity $i=1$, 
 evolves to be in cavity $i=N$, at time $t$. 
 The arrival time for perfect QST is known in certain cases,
 however, more generally, e.g. in the presence of disorder, a complication is the necessity to search for the time at
 which $f(t)$ is maximal.
 
\par It is intuitive that solutions to the time evolution equation
should usually spread in time so that the location of the quantum
particle becomes less well known. Indeed, this is also a simple
consequence of the uncertainty principle: a lack of precise knowledge
of the momentum implies that the wave packet can move with different
possible speeds and hence as time passes the distribution of possible
locations is increasingly broad. For these reasons it might appear
remarkable that there are solutions of the Schrodinger equation on a
lattice which can begin at a unique location and arrive later at a
different unique location. 
 
\par Despite this argument, it has been shown \cite{christandl2004} that, 
for a CCA with no emitters operating
in the single excitation sector,
there are a variety of prescriptions for $J_{i}$
which yield perfect QST at a known time.
For a system of $N$ cavities and $N-1$ couplings, 
one of the  simplest arrangements is: 
\begin{equation} 
J_{i} = \sqrt{i\cdot(N-i)} \, J_0
\label{eq:Jengineered}
\end{equation}
The insight here is that the hoppings $J_i$ match the Clebsch-Gordon
coefficients for the spin raising operator for spin 
$N/2-1$.  The $N$ associated eigenvalues of the $z$ component of angular momentum
are equi-spaced, allowing for a matching of phases
and hence complete re-localization of the excitation at an appropriate
future time.
Indeed, with this choice,
perfect QST occurs at $t_p = \pi/(2J_0)$ for any $N$. 
The surprising feature that the passage time is independent of chain
length $N$ is accounted for by the fact that the $J_i$ increase
with $N$. (For example, at the chain midpoint, $J_{N/2} = \frac{N}{2}$.)

\par Notice that, although we have labeled the couplings in 
Fig.~\ref{fig:cavemt} completely generally, the $J_{i}$ of Eq.~\ref{eq:Jengineered}
obey a reflection symmetry about the
chain center.  This proves to be a crucial ingredient of perfect QST\cite{yung05},
ensuring that the `return' transfer from 
$\ket{\Psi_{\cal B}}$ to
$\ket{\Psi_{\cal A}}$ 
precisely follows the transfer from
$\ket{\Psi_{\cal A}}$ to
$\ket{\Psi_{\cal B}}$.
We will reproduce these known results in the absence of emitters to provide a benchmark for our new JCHH results.

\par The geometry in the presence of emitters is shown by the 
full structure in Fig.~\ref{fig:cavemt}, i.e.~including both the cavities, represented
by the squares, and the emitters, by circles.
In this situation we will find, unsurprisingly, that the $J_{i}$
values giving perfect QST are shifted away from those of 
Eq.~\ref{eq:Jengineered}, which apply to
the cavity-only (spin chain) case.
Indeed, the discovery of a collection of $J_i, g_i$ yielding perfect QST in
the presence of emitters is one of the primary conclusions of this work.
 
\par 
Adding a single emitter to each of the $N$ cavities of our system
($\{ \, M_i =1\, \}$), but 
remaining in the one excitation
sector, the system's Hamiltonian doubles in dimension to $2N$. 
Our convention is that the first $N$ basis vectors represent photons in
cavities $i = 1,2, \cdots, N$. We acquire an additional $N$ basis vectors
$i = N+1,N+2, \cdots, 2\,N$ 
for which there are no photons but instead an emitter is excited.
The Hamiltonian matrix is now, for $N=4$,
\begin{equation} \label{eq:hamiltonian_emit}
{{\cal H} = \begin{pmatrix}
\Omega_1 & -J_{1} & 0 & 0 & -g_{1} & 0 & 0 & 0\\
-J_{1} & \Omega_2 & -J_{2} & 0 & 0 & -g_{2} & 0 & 0\\
0 & -J_{2} & \Omega_3 & -J_{3} & 0 & 0 & -g_{3} & 0\\
0 & 0 & -J_{3} & \Omega_4 & 0 & 0 & 0 & -g_{4}\\
-g_{1} & 0 & 0 & 0 & \omega_{1} & 0 & 0 & 0\\
0 & -g_{2} & 0 & 0 & 0 & \omega_{2}  & 0 & 0\\
0 & 0 & -g_{3} & 0 & 0 & 0 & \omega_{3}  & 0\\
0 & 0 & 0 & -g_{4} & 0 & 0 & 0 & \omega_{4} \\
\end{pmatrix}}
\end{equation}
This form of ${\cal H}$ has a $2 \times 2$ `block' structure reflecting the
presence of two types of `sites' in the lattice.

In the remainder of this paper, we will enforce the reflection symmetry of all
couplings in the JCHH.  That is, we will have $J_{i} =J_{N-i}$ and 
$g_{i}=g_{N-i}$.  In addition, unless otherwise stated, we set the matrix diagonals to
a common value. Since this value corresponds to the arbitrary
choice of a zero of energy, it is set to zero.  

\par While many protocols for the $J_{i}$ yielding perfect QST for the cavity only (spin chain) geometry are known, the analogous Hamiltonian parameters for perfect QST in the presence of emitters (the JCHH Hamiltonian) are, to our knowledge, not yet determined. Here we  compute appropriate couplings via a Monte Carlo procedure. We begin with the {\it assumption} that the eigenvalues for a cavity-only system of length $2N$ which give perfect QST will also give perfect QST for a JCHH system of $N$ cavities and $N$ emitters.  This starting point is motivated by the insight that the key to
perfect QST is in the (rational fraction) relation between the eigenvalues which 
allows all frequencies to be in phase at some future time. We denote these the `target' eigenvalues $\lambda_n^{(t)}$%. We
and define an action:
\begin{equation}
{\cal S} = \sum_{n} \big( \, \lambda_n - \lambda_n^{(t)} \, \big)^{2}
\end{equation}
Here $\lambda_n$ are the actual eigenvalues of the matrix ${\cal H}$ of the
JCHH Hamiltonian, Eq.~\ref{eq:hamiltonian_emit}, for a given set of
$\{ \,J_i \, \}$ and $\{\, g_i \}$.
We begin with constant $\{ \,J_i \, \}$ and $\{\, g_i \}$ and
 propose `moves' which change all the parameters within some `step size'.
We accept each move with the `heat bath' probability
 $e^{-\beta \Delta {\cal S}} \big( 1 + e^{-\beta \Delta {\cal S}} \big)^{-1} $
where $\Delta {\cal S}$ is the change in the action resulting from the Monte
 Carlo move. Here $\beta$ is a
parameter \footnote{In statistical mechanics language $\beta=1/T$ is the
inverse temperature, so that $\beta_{\rm initial}=0.1$ corresponds
to high temperature and $\beta_{\rm final}=10^4$ corresponds to low temperature.
  The gradual increase of $\beta$ (lowering of $T$) allows the
Monte Carlo to find the `ground state' ${\cal S}=0$ where the $\{ J_i \}$
and $\{ g_i \}$ give a Hamiltonian with desired target eigenvalues to high
accuracy.}
which starts at a small value (e.g.~$\beta_{\rm initial} \sim 0.1$) and
 after $L$ Monte Carlo sweeps (a typical choice was $L \sim 10^{\,6}$) of all
the parameters is increased by a factor $\alpha$. This process is
 repeated for $K$ steps until $\beta_{\rm final} = \alpha^K \beta_{\rm initial}$
 is large (e.g.~$\beta_{\rm final} = 10^4$.)

\par We find that this procedure robustly converges to small values of ${\cal S}$,
corresponding to all the eigenvalues $\lambda_n$ of ${\cal H}$ matching
their targets $\lambda_n^{(t)}$. For most results presented here we terminate
the Monte Carlo when the eigenvalues match their targets to $0.1\%$,
however, we can continue to run the program with smaller 'step sizes' 
until we reach any desired degree of accuracy. Since the fidelity of the system is dependant on the eigenvalues, this allows us to reach any desired fidelity. In this paper, we consider a fidelity of ${\cal F} \gtrsim 0.99$ as an adequate representation of perfect QST.
The time to solution scales with $N^3$ owing to the necessity of repeated
 diagonalizations of ${\cal H}$ in the computation of $\Delta {\cal S}$.
  Since our chain lengths ($N \lesssim 16$) were relatively small, the Monte Carlo time to
 solution was quite short. Such calculations can easily be done
 in a few minutes to a few hours on a desktop
 computer, depending on system size and desired accuracy \footnote{An alternate Monte Carlo procedure defines an action
 ${\cal S}$ based on targeting of a desired {\it time evolution matrix}
 rather than desired eigenvalues. 
 This procedure is useful in more complicated geometries and will be
 explored in a subsequent paper.}.
 Larger $N \sim 10^{\,2}$ are similarly quite feasible
 without resorting to specialized hardware.

\par Next we use the Hamiltonian ${\cal H}$ determined by the resulting
$\{ \, J_i, g_i \, \}$ and find that the time evolution operator 
$e^{- i {\cal H} t}$ produces {\it perfect QST for the cavity-emitter geometry.} 
This validates our assumption that the eigenvalue list is apparently what 
produces perfect QST,
 and the particular tridiagonal structure of the cavity-only (spin chain)
 matrix is not essential--it can be generalized to the $2 \times 2 $
 block matrix structure of Eq.~\ref{eq:hamiltonian_emit}
 \footnote{In future work we will explore yet more general geometries.}.

\par We note that this procedure--the computation of the matrix elements giving
a desired spectrum, or `inverse eigenvalue problem' (IEP)--is of course a well explored problem in applied mathematics
 \cite{chu2005}.
The IEP is non-trivial only when the matrix is constrained to have a 
particular structure.  The cavity-only case is that of a `Jacobi matrix' 
considered by Hald \cite{hald1976}.
Other studied structures include Toeplitz, Hessenberg, and stochastic
matrices \cite{chu2002}.
Our work addresses the IEP for an additional type of
matrix structure.

%%%%%%%%%%%%%%%%%%%%%%%%%%%%%%%%%%%
%%%%%%%%%%%%%%%%%%%%%%%%%%%%%%%%%%%
\section{QST in the Uniform JCHH}\label{sec:QST-JCHH}
%%%%%%%%%%%%%%%%%%%%%%%%%%%%%%%%%%%
%%%%%%%%%%%%%%%%%%%%%%%%%%%%%%%%%%%

\subsection{Background:  Limit of No Emitters}

\vskip0.10in
\begin{table}[t]
\renewcommand{\arraystretch}{1.2}
\hskip0.05in
\begin{tabular}{|c|c|c|}
\hline
inter-cavity &      &  \\
 bond $i$ &  $J_i\,$ (MC) & $\sqrt{i(N-i)}$ \\
\hline
 1   & 2.642   & 2.64575  \\
\hline
 2   & 3.470   & 3.46410  \\
\hline
 3   & 3.873   & 3.87298 \\
\hline
 4   & 3.996   & 4.00000 \\
\hline
 5   & 3.873   & 3.87298  \\
\hline
 6   & 3.470   & 3.46410  \\
\hline
 7   & 2.642   & 2.64575 \\
\hline
\end{tabular}
\caption{Values of the cavity-only (spin chain) couplings determined by the 
Monte Carlo for a $N=8$ compared to the known results for perfect QST
given by Eq.~\ref{eq:Jengineered}.
Our Monte Carlo enforces the symmetry $J_i = J_{N-i}$.
}
\label{table:JCHHJ}
\end{table}

\begin{figure}[t] 
\includegraphics[width=1\columnwidth]{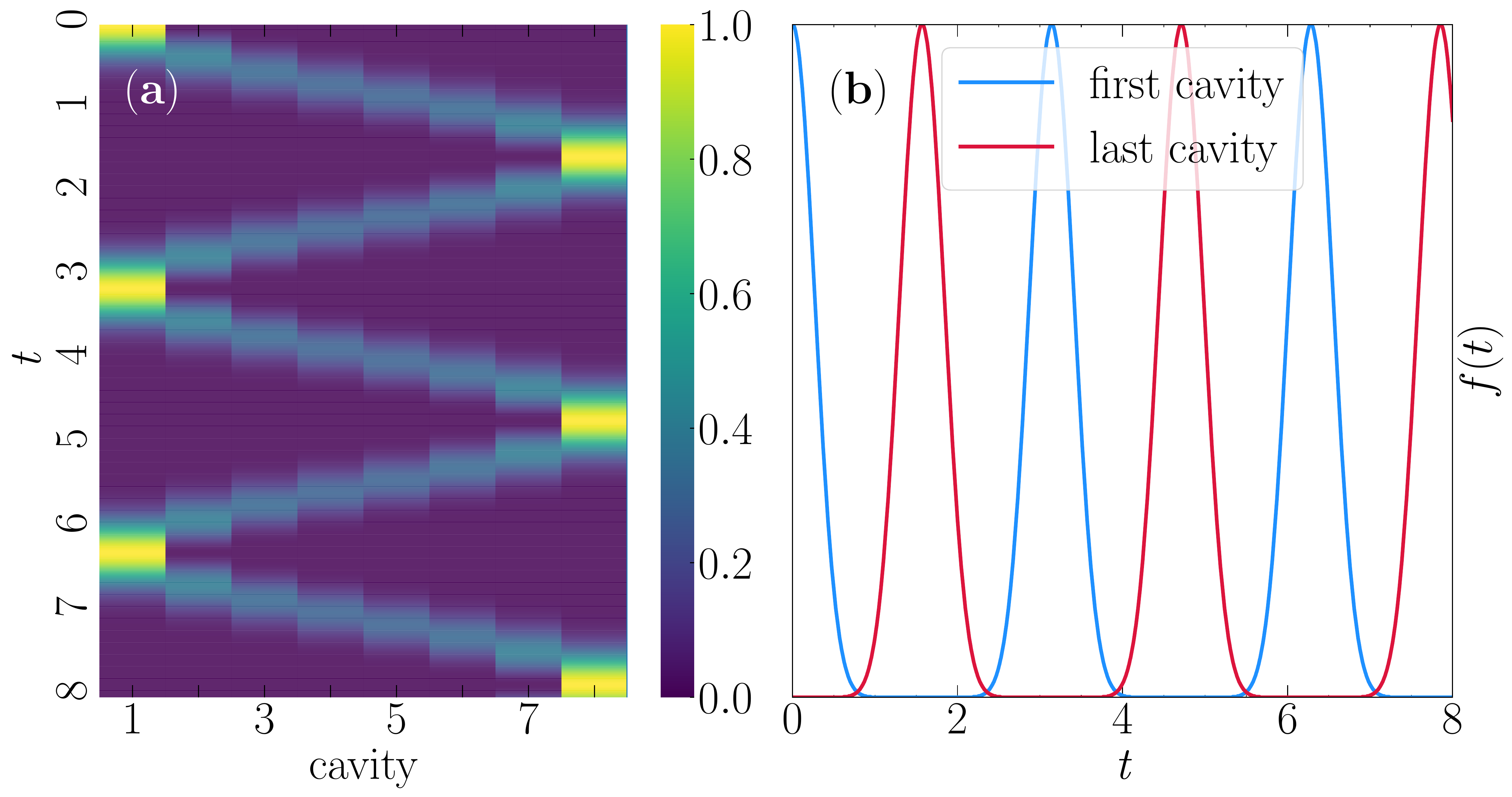}
\caption{We consider an $N=8$ cavity only system with all $\Omega_i = 0$ 
and $J_i$ couplings determined by Monte Carlo in \ref{table:JCHHJ}, which
converges to the values of the analytical solution. In panel 
\textbf{(a)} we graph the probability that the photon is in each cavity.
The eight columns on the x axis represent the eight
 cavities, and time descends from 0 to 8 along the y axis. For every
 time and location, the probability is 
 indicated in the color bar. In panel \textbf{(b)} we display the probabilities
 in just the originating and receiving cavities as functions of time.
 We observe perfect QST at time $\pi/2$ and with period $\pi$ for a return
 to the initial state.
}
\label{fig:QST8}
\end{figure}

Here we reproduce the known results of Christandl \cite{christandl2004} 
in the absence of randomness to serve as a point of comparison for our
 subsequent
study of the JCHH, and to test our Monte Carlo method for
the IEP in a situation where a solution is already established. 
We therefore consider a cavity-only system with near neighbor couplings.
%% and proceed with our Monte Carlo methodology.  
We confirm rapid and precise convergence to the known perfect QST 
values of 
 Eq.~\ref{eq:Jengineered} from general, random starting configurations
 of $\{ J_i \}$. We compare our results in 
 Table \ref{table:JCHHJ} to the exact values of 
Eq.~\ref{eq:Jengineered} for $N=8$ and 
target eigenvalues $\lambda_n^{(t)} =
 \pm \frac{1}{2}, \pm \frac{3}{2}$.
 Figure \ref{fig:QST8} gives the resulting time evolution.
 The heat map of the left-hand panel displays the probability in each cavity 
for all times.  We supplement
 this (right-hand panel) with a fidelity line graph for the first and last cavities, 
where the probabilities can be displayed more precisely. 
The small deviations of $J_i$ from the analytic values do not appreciably
degrade the fidelity.

\subsection{QST in the Presence of Emitters}

\par Next we demonstrate the effectiveness of our Monte Carlo solution 
of the IEP for determining cavity-cavity and cavity-emitter coupling leading to
perfect QST in the novel context of  the  JCHH.  
Our method works only
 with systems with an even number of cavities when there is
 one emitter in every cavity.  The reason is discussed further in the Appendix.
 However, with this constraint, we can successfully determine JCHH parameters
 giving fidelities ${\cal F} \gtrsim 0.99$ for
systems with up to $N \sim 10^{\,2}$ cavities. 

\par
\begin{figure}[t] 
\includegraphics[width=1\columnwidth]{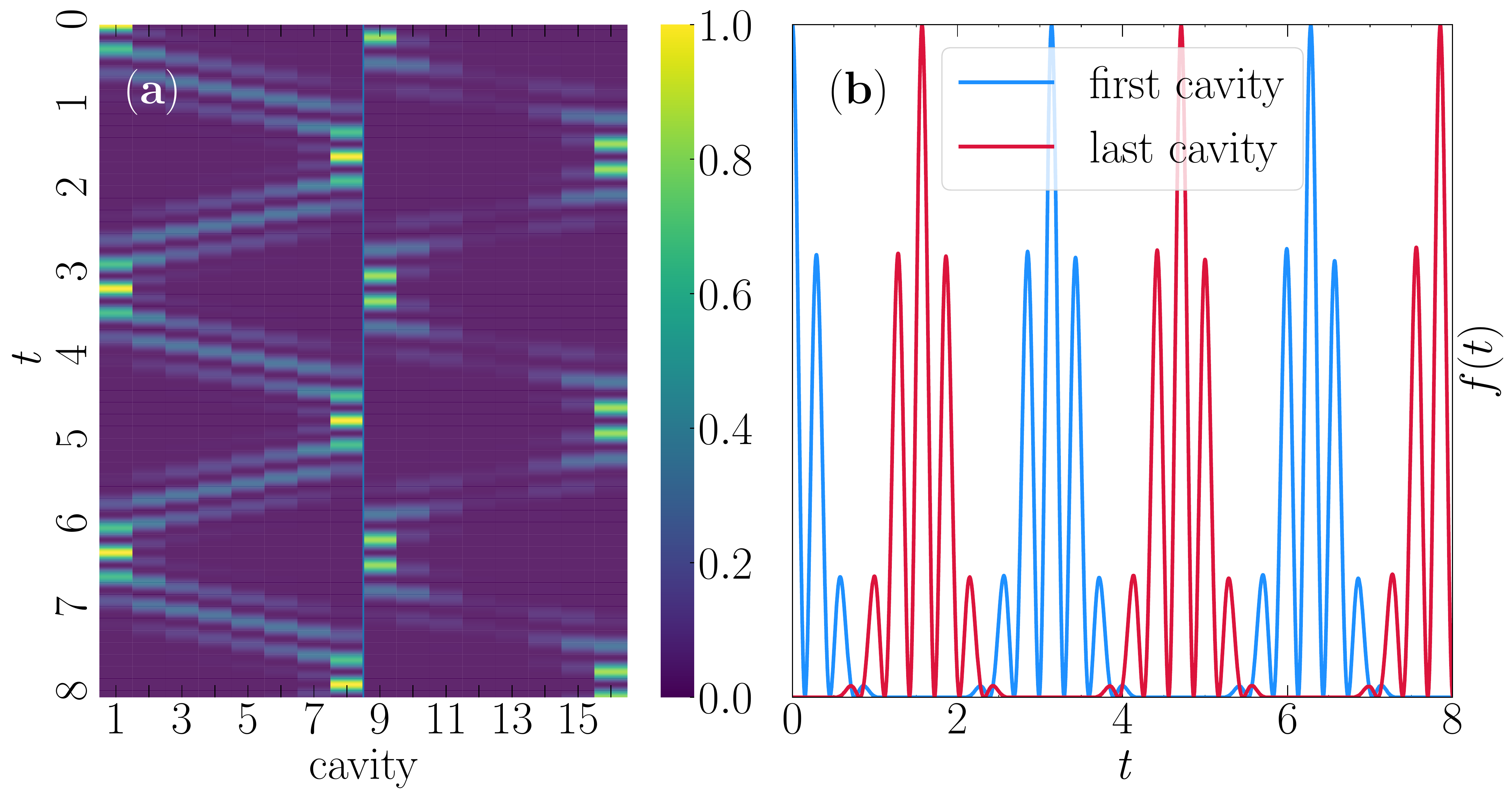}
\caption{We consider a JCHH system of eight cavities and eight emitters
with all $\Omega = \omega = 0$ and couplings according to \ref{table:JCHHJg}. 
Target eigenvalues were chosen to be those giving perfect QST for
a $N=16$ cavity-only chain, i.e.~using  Eq.~\ref{eq:Jengineered} with $N=16$.
In panel
 \textbf{(a)} we graph the probability that the photon is in each cavity and
 in each emitter for multiple times. The sixteen columns on the x axis
 represent the eight cavities and eight emitters, and time descends
 from 0 to 8 along the y axis. For every time and location, the
 probability is 
 indicated in the color
 bar. In panel \textbf{(b)} we display the probability in starting and
 receiving cavities as a function of time. We observe perfect
 QST at time $\pi/2$ and with period $\pi$.}
\label{fig:QST88}
\end{figure}

 Perfect QST for a system of eight cavities with emitters in every cavity 
is shown in Fig.~\ref{fig:QST88}.  Our labeling convention is such that
we index of states with a photon in one of the $N=8$ optical 
cavities as 1-8, and states with the corresponding emitter in
an excited level as 9-16. 
As with Fig.~\ref{fig:QST8},
the left panel is the heat map of the probability in all `sites' (cavities {\it and} emitters), 
whereas the right panel focuses on the originating and receiving cavities only.
We see that perfect QST is obtained 
in this `8+8' JCHH system.
However,
the time evolution is considerably more complex than for the cavity-only 
(spin) system of 
Fig.~\ref{fig:QST8}.
The transfer time remains $\pi/2$, but the peaks now form envelopes containing
an additional higher frequency structure.
This results from a rapid transfer of probability between each cavity and its
associated emitter
which occurs as the overall probability moves, with a longer
time scale, down the cavity backbone.

Table \ref{table:JCHHJg} gives the values of the JCHH Hamiltonian
parameters determined by our Monte Carlo and yielding the time evolution of
Fig.~\ref{fig:QST88}.
Values for $J_i$ and $g_i$ for several other $N$ are given in the Appendix,
as is a discussion of an empirical formula which gives a reasonable
fit to the data.

\vskip0.10in
\begin{table}[t]
\renewcommand{\arraystretch}{1.2}
\hskip0.05in
\begin{tabular}{|c|c|}
\hline
inter-cavity &       \\
 bond $i$ &  $J_i\,$  \\
\hline
 1   &  4.521  \\
\hline
 2   &  6.158  \\
\hline
 3   &  7.232  \\
\hline
 4   &  7.979  \\
\hline
 5   &  7.232  \\
\hline
 6   &  6.158  \\
\hline
 7   &  4.521  \\
\hline
\end{tabular}
\hskip0.25in
\begin{tabular}{|c|c|}
\hline
cavity-emitter &       \\
bond $i$ &  $g_i\,$  \\
\hline
 1   &  9.558  \\
\hline
 2   &  7.825  \\
\hline
 3   &  5.872  \\
\hline
 4   &  3.234  \\
\hline
 5   &  3.234  \\
\hline
 6   &  5.872  \\
\hline
 7   &  7.825  \\
\hline
 8   &  9.558  \\
\hline
\end{tabular}
\caption{Values of the JCHH couplings determined by the Monte Carlo for
an $N=8$ cavity array with an emitter in each cavity.  Note that
bond $J_i$ connects cavities $i$ and $i+1$ whereas
bond $g_i$ connects cavity $i$ with its associated emitter. See
Fig.~\ref{fig:cavemt}.
}
\label{table:JCHHJg}
\end{table}

\vskip0.10in \noindent

%%%%%%%%%%%%%%%%%%%%%%%%%%%%%%%%%%%%%%
%%%%%%%%%%%%%%%%%%%%%%%%%%%%%%%%%%%%%%
\section{Effect of Emitters on Perfect cavity-only QST}\label{sec:QST-JCHH-dis}
%%%%%%%%%%%%%%%%%%%%%%%%%%%%%%%%%%%%%%
%%%%%%%%%%%%%%%%%%%%%%%%%%%%%%%%%%%%%%

In the preceding section we demonstrated that perfect QST is
possible for systems with uniform arrangements of emitters,
precisely one per cavity.  We now consider a distinct issue,
namely what effect a single `impurity' emitter would have
on the perfect QST which would occur in a cavity-only system.
This explores a different type of `disorder' from that
considered previously, and is experimentally relevant, since
in cavity-emitter systems fluctuations in the numbers of emitters in each cavity are to be expected.

\subsection{Background:  Limit of No Emitters}

\begin{figure}[t] 
\includegraphics[width=1\columnwidth]{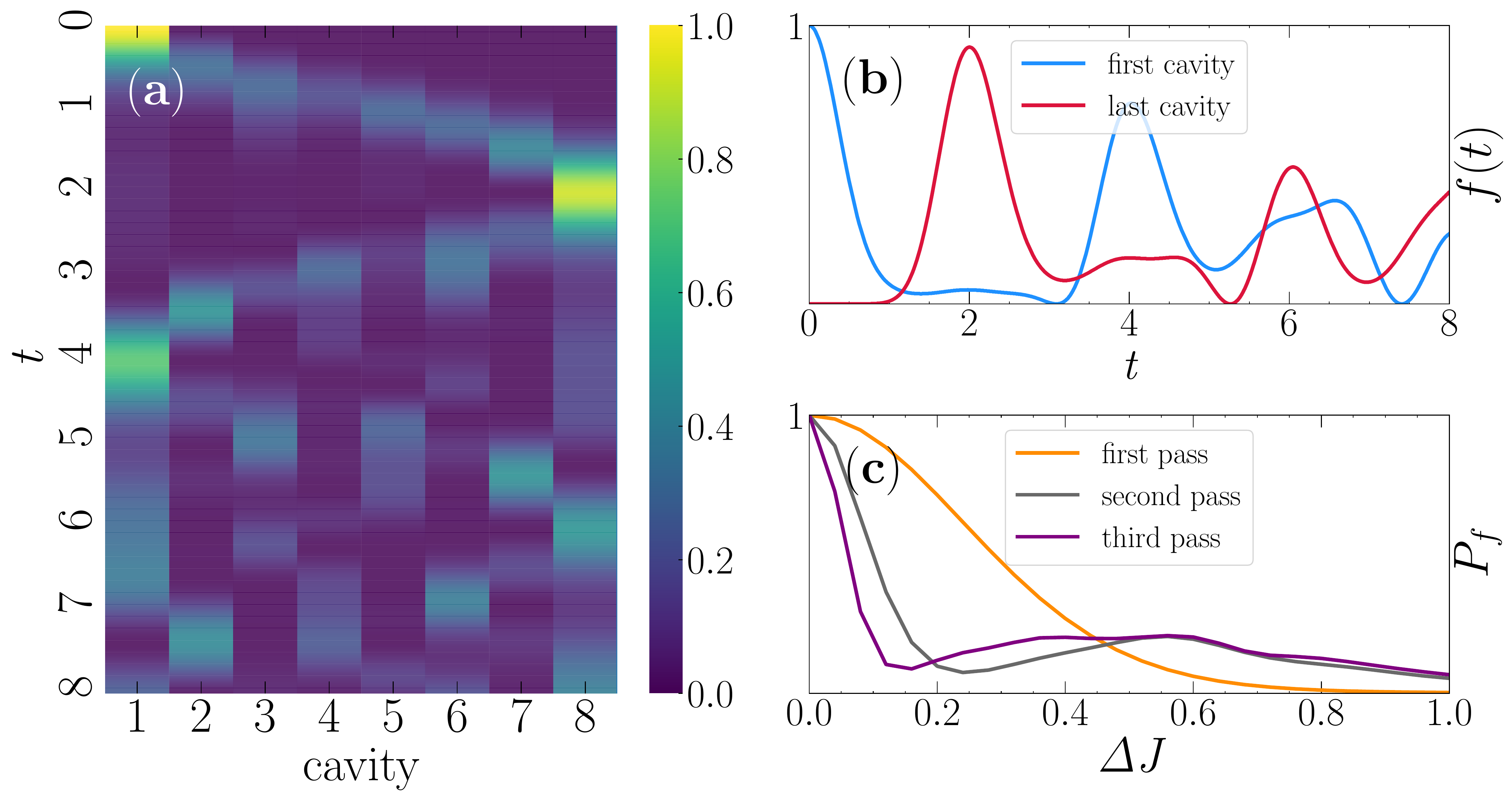}
\caption{State transfer in a system of eight cavities with 
 couplings according to Equation \ref{eq:Jengineered}. A random disorder
 between $-\Delta J$ and $+\Delta J$ is added to each $J_i$.
\textbf{(a):} Probability heat map 
for a single realization with  $\Delta J= 0.5$.
\textbf{(b):}
Associated fidelity line graph.
 We observe that perfect QST does not occur, and the fidelity in the
 last cavity decreases after each ``pass." 
\textbf{(c):} we
 graph the average fidelity for an $N=8$ cavity perfect QST system with
 varying levels of disorder $\Delta J$. The fidelity is measured at the
 expected transfer time $\frac{\pi}{2}$, and the average is taken over
 $10^{\, 4}$ disorder realizations.
 }
\label{fig:QSTHoppingDisorderCombined}
\end{figure}

Again, we begin by establishing context for our new results on 
the effect of disorder in the JCHH by re-examining the cavity-only
system previously considered in \cite{zwick11,zwick15} .
We set $J_0=1$ as our scale of energy (time$^{-1}$) and add an `absolute' random noise
of scale $\Delta J=0.5$ to each of the engineered 
$J_i$ \footnote{Situations where the randomness is `relative', 
i.e.~scaled to the $J_i$ on each site, have also been studied \cite{zwick15}}. 
We observe in Fig.  \ref{fig:QSTHoppingDisorderCombined} that,
 while we still see the oscillations present in the
 perfect system, the added noise significantly degrades QST.

By calculating the fidelity at $t = \frac{\pi}{2}$ for many values of
 $\Delta J$ and taking the average fidelity over $10^{\,4}$
 realizations of randomized
 disorder, we can determine the effect $\Delta J$ has on the fidelity. 
 To emphasize the distinction from the fidelity for the clean system
 or for a single realization,
 we denote this average as ${\cal P}_f$.  We
 obtain ${\cal P}_f$ for the first, second and third passes, where the $n$th pass
 is the fidelity taken at $t_n = \frac{\pi}{2} + (n-1)\cdot \pi$. The results
are displayed in the bottom right panel of Fig.~\ref{fig:QSTHoppingDisorderCombined}.
 The fidelity at the first pass decreases as the disorder
 increases, and in each successive pass the fidelity decreases more
 steeply. The second and third passes undergo a
 small rise after their initial declines, but this quickly flattens out. 
 This non-monotonicity with $\Delta J$ 
is associated with the way in which the data are extracted:
we measure $f(t)$ for each realization at the fixed 
clean system transfer time $t_n$.
 However $\Delta J$ not only disrupts the phase matching of the engineered
 $J_i$, it also alters the speed of propagation.  An alternate 
 (and computationally time-consuming) protocol would be to search over time for 
 the optimal fidelity for each $\Delta J$ and for each realization.  

\begin{figure}[t] 
\includegraphics[width=1\columnwidth]{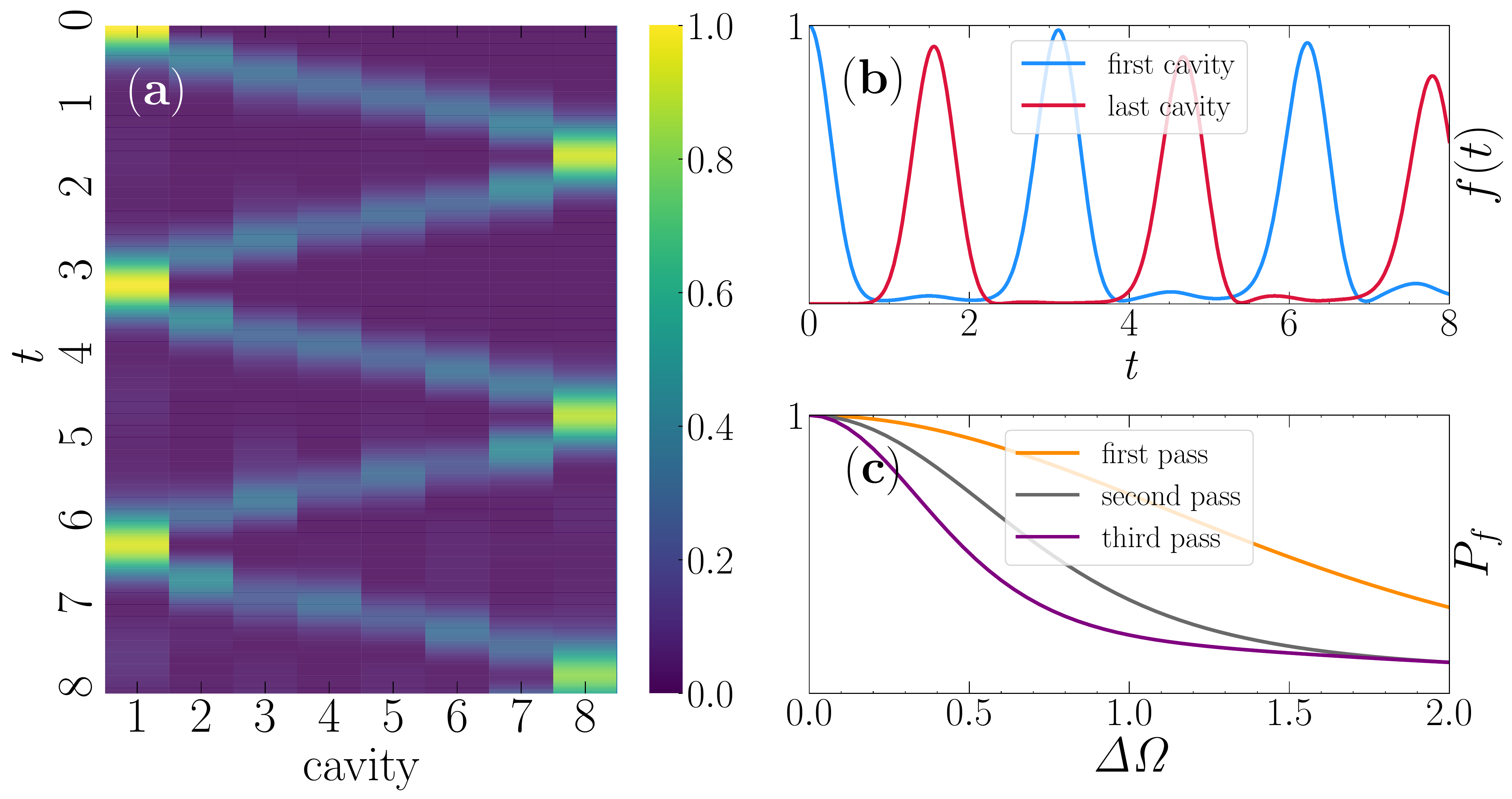}
\caption{
Analog of Fig.~\ref{fig:QSTHoppingDisorderCombined}
except for disordered cavity energies rather than inter-cavity hopping.  
panels \textbf{(a)} and \textbf{(b)} give results for a single realization with 
$\Delta \Omega=1$, while panel \textbf{(c)} shows averages over many realizations
for different $\Delta \Omega$.  The fidelity loss appears to be roughly linear in 
the pass number for small $\Delta \Omega$; that is, the deviation in the maxima in the fidelity
grow roughly linearly with $n$.
 }
\label{fig:QSTDisorderEnergyExample}
\end{figure}

We can also quantify the effects of random $\Omega_i$ by adding
noise so that the cavity energy levels are uniformly
distributed on $\big(-\frac{\Delta \Omega}{2}, +\frac{\Delta \Omega}{2}\big) $. 
Such randomness can arise from variations in the size and shape of the cavities.
Our observations
(Fig.~\ref{fig:QSTDisorderEnergyExample})
 are similar to our discussion of hopping disorder: We see oscillations
 with peaks that successively decline. 

We now turn to analyzing the effects of adding atom-like emitters to our
cavity-only system.  We will first
consider the effect of
 adding a single emitter to a cavity-only geometry
 with $J_i$ engineered to give perfect QST. 
We will next consider cases with many periodically placed, but
non-uniform, emitters (random $g_i$ and $\omega_i$).
 The subsections below
 analyze these two situations. 

\subsection{Loss of Fidelity due to Emitters}

\begin{figure}[t] 
\includegraphics[width=1\columnwidth]{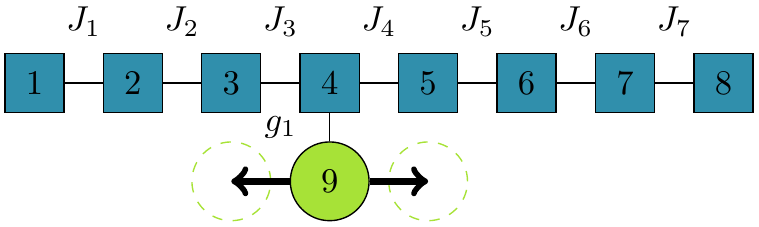}
\caption{
Geometry of the one dimensional \tcb{(sparse)} Jaynes-Cummings-Hubbard Hamiltonian
with a single emitter.  
We will focus on the emitter's effect on the fidelity for
values of $J_i$ which give perfect QST for a cavity-only
system.
}
\label{fig:cav1emt}
\end{figure}

\begin{figure}[t] 
\includegraphics[width=1\columnwidth]{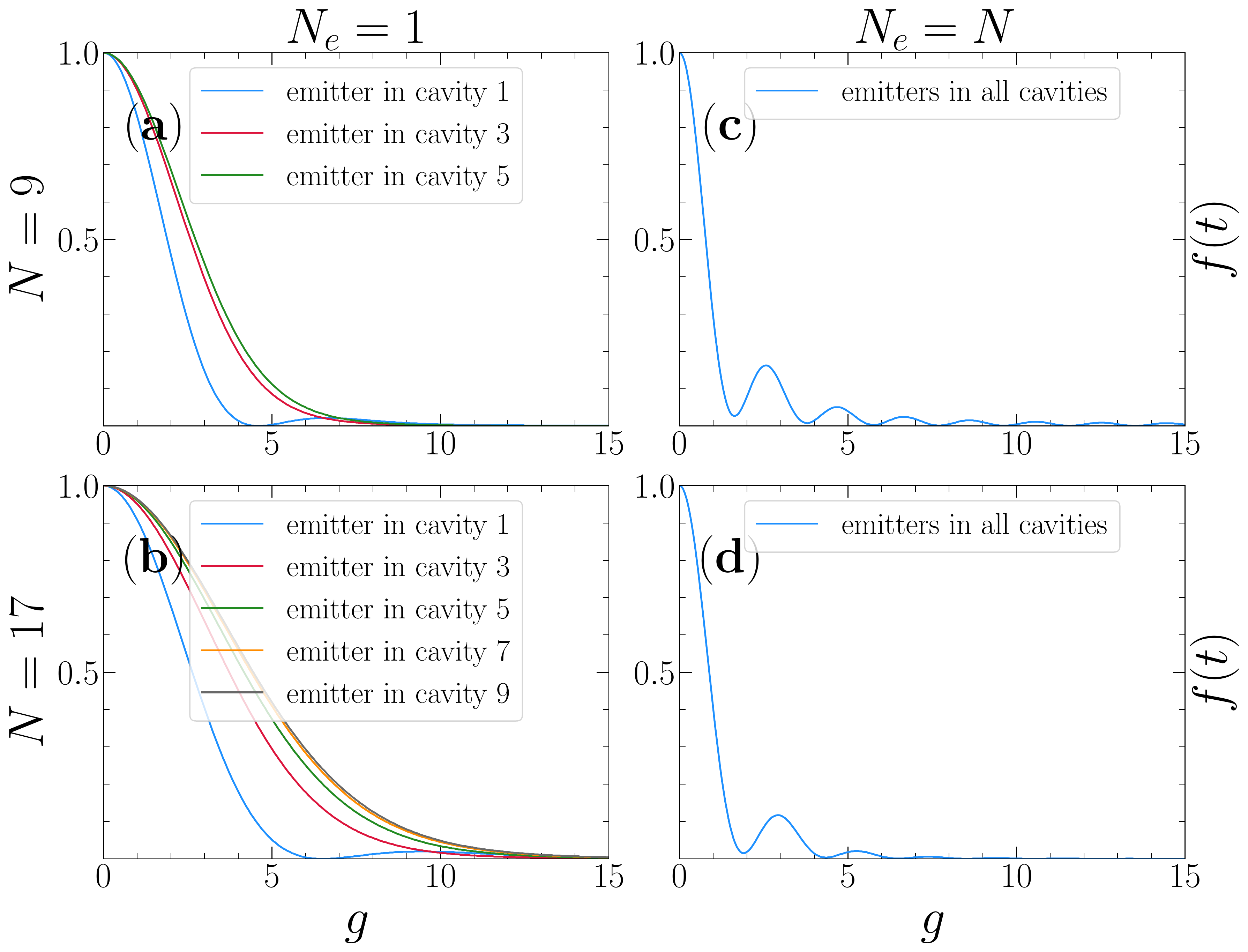}
\caption{Effect of the addition of a {\it single} emitter as 
a perturbation to cavity-only perfect QST. 
\textbf{(a):} Fidelity $f(t=\pi/2)$ as a function of the coupling $g$
of the (single) `impurity' emitter to its cavity.  Curves for
three emitter placements, cavities 1,3,5 are shown.
(Reflection symmetry implies the effect of an emitter
in cavity $N-i$ is identical to that of an emitter
in cavity $i$.)
The number of cavities $N=9$.
\textbf{(b):}  Same as \textbf{(a)} except for $N=17$.
\textbf{(c):}  Fidelity $f(t=\pi/2)$ as a function the coupling $g$
of a collection of emitters, one in each cavity,
as a perturbation to cavity-only perfect QST.  
The number of cavities $N=9$.
\textbf{(d):}  Same as \textbf{(c)} except for $N=17$.
 }
\label{fig:EmitterAsDisorder}
\end{figure}

Figure \ref{fig:cav1emt} shows the first geometry we consider:  a {\it single
emitter} is added to a chain of  $N$ cavities with couplings $J_i$.
The position of the emitter is variable.  
The left panels of Fig.~\ref{fig:EmitterAsDisorder}
describe the effects of such an impurity emitter 
on a cavity system with $J_i$ engineered to perfect QST.
Results for different emitter cavity coupling $g$ and
emitter placement are shown.  An emitter at the edge
of the chain (i.e.~close to either the origin cavity
or the destination cavity) causes the most rapid
fidelity loss.  It is interesting
that the disruption of QST is less severe 
as the chain length increases
(bottom left compared to top left).
As with the independence of passage time on $N$, it is possible
this greater robustness of perfect QST with $N$ is associated
with the increasing values of $J_i$.

The right panels of Figure \ref{fig:EmitterAsDisorder}
consider another type of emitter disruption, namely a situation
where an emitter is present in each cavity (all with the
same $g_i=g$). 
The fidelity falls more rapidly with $g$ than
for a single emitter (left panels), but there are periodic
fidelity `revivals' which are associated with the more regular
geometric structure of uniform emitter placement.

\begin{figure}[t] 
 \includegraphics[width=0.6\columnwidth]{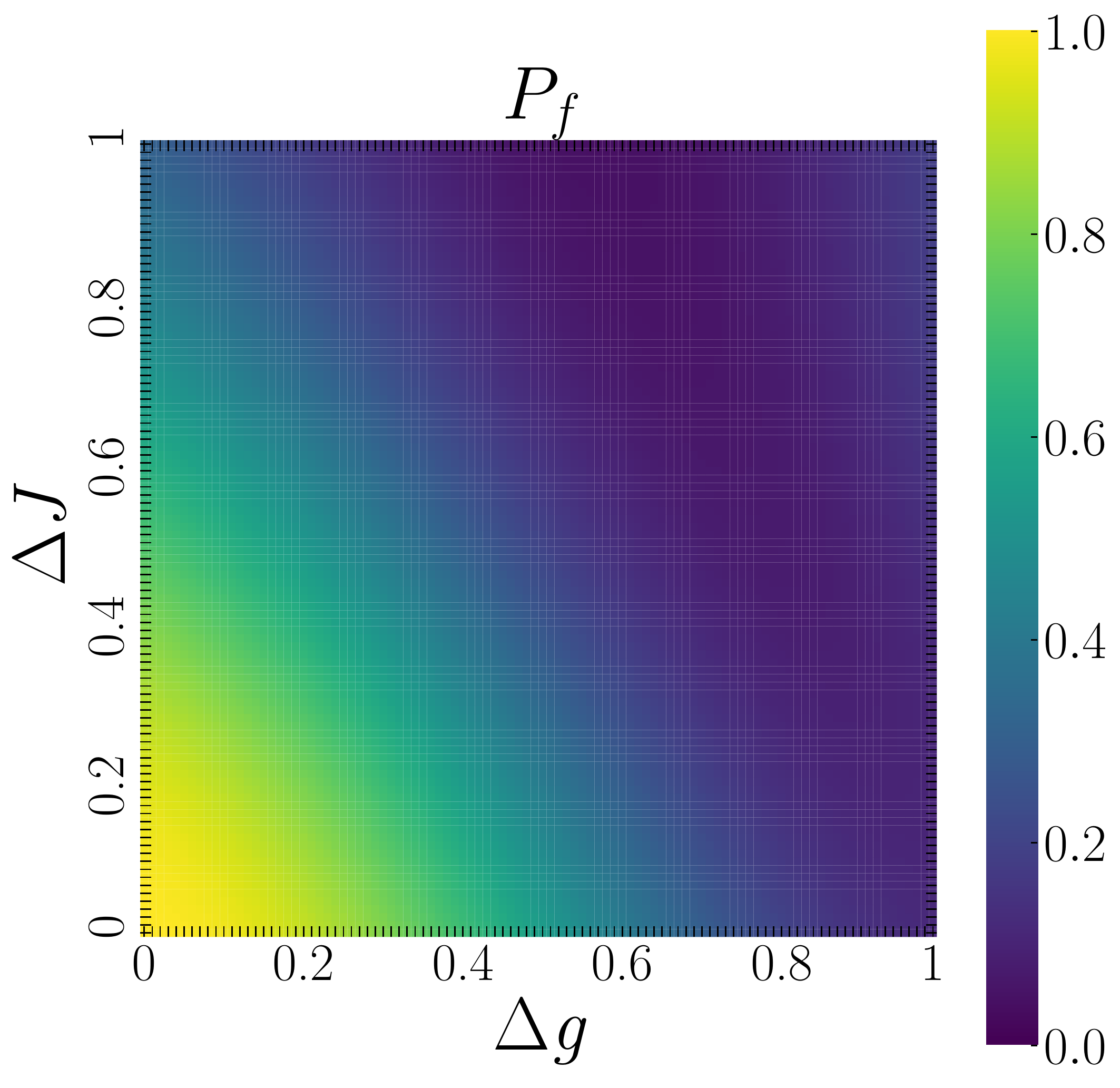}
\vskip0.10in \noindent
\caption{
Fidelity $f(t=\pi/2)$ in the JCHH system of eight cavities and eight emitters as a function of varied cavity and emitter coupling disorder. The initial couplings are determined by the Monte Carlo procedure according to \ref{table:JCHHJg}. For each realization, every coupling $J_i$ is randomly disordered between $J_i \pm \frac{\Delta J}{2}$. Similarly, each $g_i$ coupling is randomly disordered between $g_i \pm \frac{\Delta g}{2}$. Each data point is taken as the average fidelity over 200 realizations.
}
\label{fig:Fheatmap}
\end{figure}

Finally, we examine disorder which has a similar form
to randomness in $J_i$ considered in earlier spin-chain 
studies \cite{zwick11}.
Specifically, we consider a situation of $N$ cavities, each with
an emitter, but allow both the intercavity hoppings
to be random on $(J_i-\frac{\Delta J}{2}, J_i+\frac{\Delta J}{2})$, 
and the emitter-cavity couplings
to be random on $(g_i-\frac{\Delta g}{2}, g_i+\frac{\Delta g}{2})$, with
$J_i$ and $g_i$ according to Table \ref{table:JCHHJg}.
The heat map of Fig.~\ref{fig:Fheatmap}
gives the realization-averaged fidelity 
$P_f(\Delta g,\Delta J)$.  The deterioration
of perfect QST is more rapid here than in 
Fig.~\ref{fig:EmitterAsDisorder} because we not only
have 
additional transfer
paths provided by the emitters, but also these paths
themselves have randomized hopping.

%%%%%%%%%%%%%%%%%%%%%%%%%%%%%%%%%%%%
%%%%%%%%%%%%%%%%%%%%%%%%%%%%%%%%%%%%
\section{The JCHH as a Realization of Boundary Engineering}\label{sec:JCHHBE}
%%%%%%%%%%%%%%%%%%%%%%%%%%%%%%%%%%%%
%%%%%%%%%%%%%%%%%%%%%%%%%%%%%%%%%%%%

This short section mainly makes an observation about
an intriguing connection between `boundary engineering'
commonly discussed in spin chains \cite{banchi2013}
and cavity emitter systems.  Topologically, 
and in the single excitation sector, a single emitter
in an end cavity behaves identically to an additional cavity
with $g$ playing the role of $J$.  Thus there is a precise
equivalence between the Hamitonian matrix and hence QST of 
systems with $N-2$ cavities and two `end' emitters and
ones with $N$ cavities and no emitters.

This mapping is especially interesting in that the 
known prescription for good QST when the $J_i$ are uniform
except at the end requires $J_1$ and $J_N$ to be much less
than the other, uniform $J_i$ in the chain interior.
Such a situation arises very naturally in cavity-emitter systems.
Hence this might be a promising alternate way to construct 
boundary engineered systems.

\begin{figure}[t] 
\includegraphics[width=1\columnwidth]{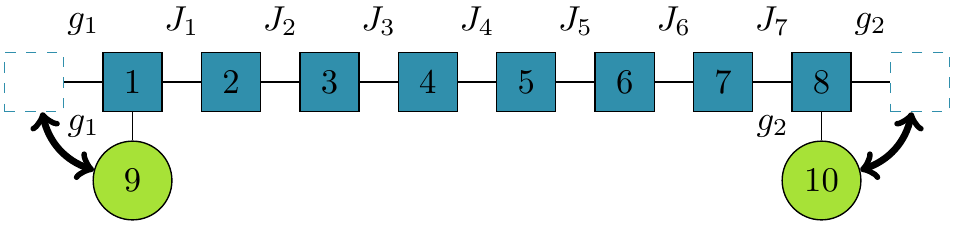}
\caption{
Geometry of the one dimensional Tavis-Cummings-Hubbard Hamiltonian with emitters
only in the first and last cavities.  The arrows indicate the geometric
equivalence of this structure to a `boundary engineered' spin chain
for which only the first and last exchange constants $J_{1}=J_{7}=g$
are different from the bulk value $J$.
}
\label{fig:cavemtbe}
\end{figure}

%%%%%%%%%%%%%%%%%%%%%%%%%%%%%%%%%%
%%%%%%%%%%%%%%%%%%%%%%%%%%%%%%%%%%
\section{Multiple Excitations}\label{sec:Multi}
%%%%%%%%%%%%%%%%%%%%%%%%%%%%%%%%%%
%%%%%%%%%%%%%%%%%%%%%%%%%%%%%%%%%%

Coupled cavity arrays differ from spin chains if more than one photon is
present in the array, since two photons can occupy the same cavity, whereas
an emitter can only be excited a single time.
A final avenue of investigation described here concerns the case of 
multiple excitations in cavity-emitter systems.   
In the absence of emitters, perfect QST in the single excitation sector automatically
implies perfect QST for multiple excitations:  the photons are {\it non-interacting}
particles.  When emitters are present, this theorem no longer holds:  
emitters can only be excited once and hence the two excitation
sector differs in a fundamental way from single excitation sector.
Another way to phrase the non-triviality of multiple excitations
 is to note
that even though the Hamiltonian is quadratic in the creation and destruction
operators, usually a hallmark of the absence of interactions, the {\it mixed}
nature of the allowed occupations introduces an effective
`many-body' correlation between excitations,
in the sense that the eigen-energies of the two particle system are not sums of
the single particle eigen-energies, as they would be if the character of the
operators were purely bosonic or purely fermionic.
Zhu {\it etal} have considered the contact interaction 
induced by the non-linearity of the JCHH in the context of
the two-polariton scattering problem\cite{zhu13}.

\begin{figure}[t] 
\includegraphics[width=1\columnwidth]{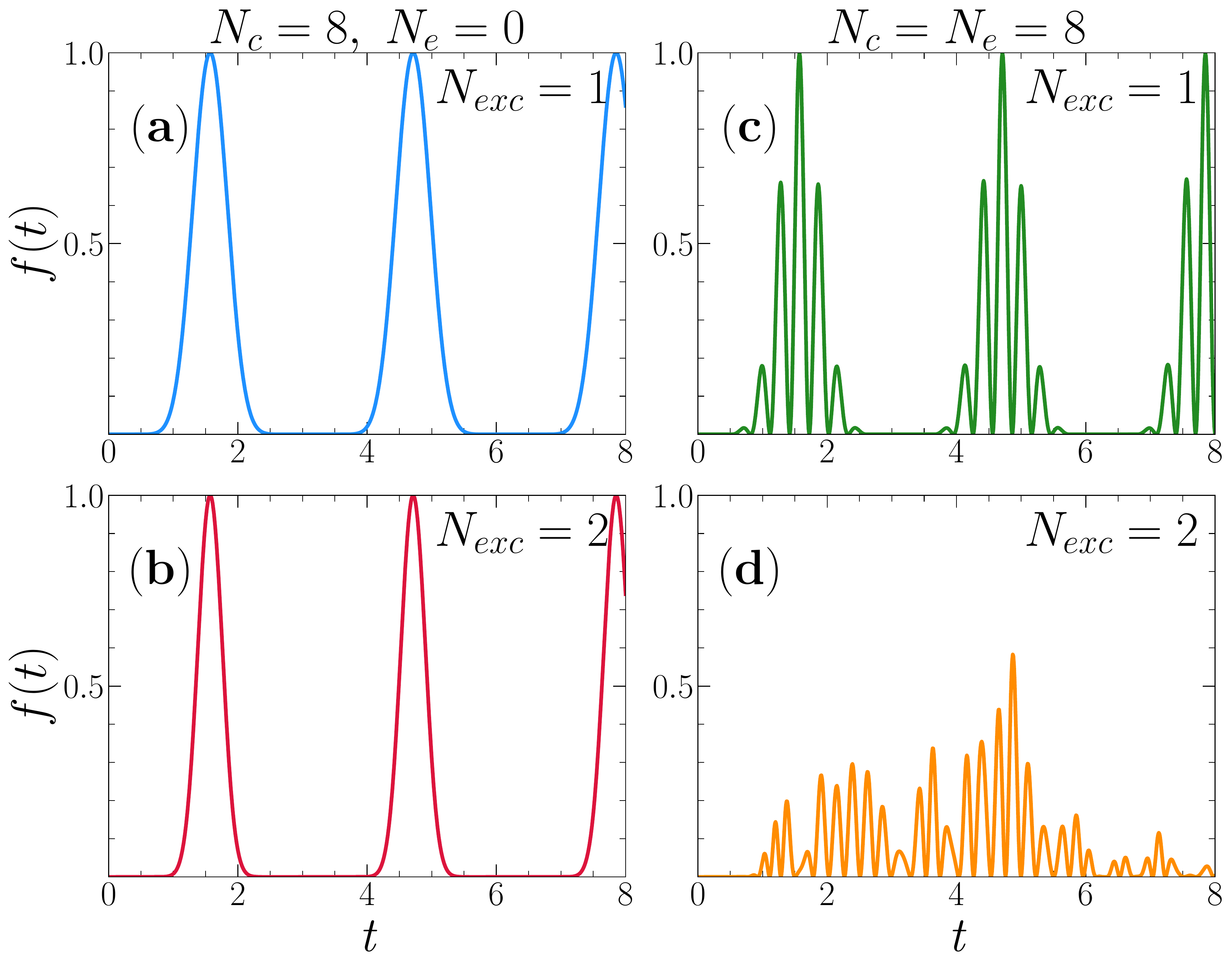}
\caption{In the left column, the fidelity of a cavity-only system
with a single excitation (top) and two excitations (bottom).
The hoppings $\{ \, J_i \, \}$ that give perfect QST for
$N_{\rm exc}=1$ also do so for  $N_{\rm exc}=2$, reflecting the non-interacting nature of the system.
In the right column, similar data for the
 cavity-emitter system as shown in Fig.~\ref{fig:cavemt},
 with one excitation (top) and multiple excitations (bottom).
 For the cavity-emitter system, 
the JCHH parmeters $\{ \, J_i,g_i \, \}$ (determined by Monte Carlo) that give perfect QST for
$N_{\rm exc}=1$ fail to give good fidelity for $N_{\rm exc}=2$.  This is a consequence of the mixed bosonic/fermionic character of the JCHH operators which
comes into play when $N_{\rm exc} > 1$.
 }
\label{fig:QSTJCHHmulti}
\end{figure}

Figure \ref{fig:QSTJCHHmulti} makes this observation more precise.  The left panels are for a cavity only system
with a single excitation at top, and two excitations at bottom.
The same $\{ \, J_i \, \}$ are used in the two cases.
Perfect QST is preserved for multiple excitations\cite{perez13}.  
The only difference is that the arrival time is more narrowly
defined for two excitations.

On the other hand, in the two right panels, which
are for a cavity-emitter system, perfect QST occurs
in the case of a single excitation but is destroyed in the case of two excitations. As with the cavity-only geometry
our procedure is to find the 
$\{\, J_i, g_i \, \}$ which work for a single excitation 
(by targeting eigenvalues for a $2\,N$ cavity-only system
as discussed earlier) and
then simulate what happens for two excitations.
We conclude that
the `effective interaction' induced by the
mixed commutation rules introduces inter-particle scattering
during the propagation.

A possible way to recover perfect QST for multiple excitations
in the cavity-emitter case would be to use a different set of couplings
for two excitations than for one.
However, finding such a set is not straightforward.
For single excitation systems, $N$ devices (cavities and emitters) will
always have $N$ basis states, thus, given any given configuration of cavities
and emitters, there exists a cavity only system with the same number of
basis states. This means you can always tune these systems as you can
create perfect QST systems with the same number of eigenvalues. This ceases
to be the case for more than one excitation.  

\vskip0.10in \noindent

%%%%%%%%%%%%%%%%%%%%%%%%%%%%%%%%%%
%%%%%%%%%%%%%%%%%%%%%%%%%%%%%%%%%%
\section{Experimental Parameters}\label{sec:exptparam}
%%%%%%%%%%%%%%%%%%%%%%%%%%%%%%%%%%
%%%%%%%%%%%%%%%%%%%%%%%%%%%%%%%%%%

We discuss here the typical range of values for
the parameters in the JCHH which would arise 
in one of its potential realizations.

The proposed coupled cavity arrays with quantum emitters are well suited for implementations in color center platforms, such as silicon carbide and diamond. Color centers are quasi-atoms formed within the lattice defects of a semiconductor emitting at visible and near infra-red frequencies, 200 THz $< \omega/2\pi < $500 THz \cite{norman21}.
Recently, significant progress has been made in fabrication of optical cavities in these materials ($\Omega \approx \omega$) and the engineering of light and matter interaction with  rates of $g/2\pi \sim$ 5 GHz \cite{evans18}.
This level of interaction, several orders of magnitude higher than achievable in atomic cavity QED systems, is a consequence of the large dipole momentum of color centers and the small mode volume of the cavities. It is worth noting that the optimal positioning of the color center, resulting in maximal $g$ value, is at the maximum of the electromagnetic field of the optical mode. An ensemble integrated into the cavity is likely to have a variation in individual emitter-cavity coupling rates. Scaling these systems into an array, photonic designs of coupled cavities have been proposed for a range of hopping rates 1 GHz $<J/2\pi<$ 200 GHz  \cite{majety21}.
Variation of nanofabrication conditions across the sample may cause a variation in resonant frequencies of each cavity, however, methods such as photo-oxidation  \cite{piggott14}.
can be used to shift resonances and synchronize the system. Finally, intrinsic as well as fabrication-induced strain in the sample causes spectral disorder among color centers. This inhomogeneity has typically been in the $\sim$10 GHz range for a variety of emitters in silicon carbide and diamond \cite{schroder17,babin21}.%

A link between fluctuations in $g_i$ and in emitter locations
is that in a cavity with $M_i$ emitters there is a 
renormalization of the emitter-cavity coupling 
$g \rightarrow g\sqrt{M}$, or more specifically $\sqrt{\sum_{j=1}^M g_j^2}$) to form a polariton state. 
Thus fluctuations in $\{\,M_i\,\}$ serve as an additional
source of  randomness in $g_i$.

%%%%%%%%%%%%%%%%%%%%%%%%%%%%%%%%%%
%%%%%%%%%%%%%%%%%%%%%%%%%%%%%%%%%%
\section{Conclusions}\label{sec:Conclusions}
%%%%%%%%%%%%%%%%%%%%%%%%%%%%%%%%%%
%%%%%%%%%%%%%%%%%%%%%%%%%%%%%%%%%%

Over the past two
decades, the experimental realization of individual optical cavities, and their
assembly into a CCA \cite{saxena2021photonic},
has allowed for the study of a wealth of quantum many-body phenomena,
including the simulation of strong correlation phenomena encountered in 
condensed matter physics \cite{hartmann08, smith2021exact}. 
As with their ultracold atom, optical lattice 
counterparts \cite{schafer20,esslinger10},
cavity QED systems permit the manipulation of 
individual system components. This level of experimental control
makes them attractive candidates for performing simulations
of superfluid to Mott insulating behavior, Anderson localization, {\it etc}.
When emitters are also present, new effects occur, 
including the emergence of polaritons, or quasiparticles
consisting of a superposition of photonic and atomic excitations \cite{bose07b,hartmann08,almeida16,hartmann06}.  The study of polaritons allows 
new strongly correlated regimes of light-matter interaction to be probed.  Our study of quantum state transfer in such systems is complementary
to those endeavors.

There are interesting analogies between the geometry considered here
and that of the one dimensional Kondo or Periodic Anderson Hamiltonians.
In those canons of condensed matter physics, 
electron hopping occurs between sites of a 
`conduction band' 
(hence the analog of cavities here)
while there are also `localized electrons' 
which hybridize with their conduction electron
partners but not each other (the analogs of emitters).  The single particle physics of the 
periodic Anderson Hamiltonian is well understood:
a hybridization gap opens where the flat impurity
band crosses the conduction band.
Our work directly connects to the QST problem
in a one-dimensional, non-interacting, 
periodic Anderson Hamiltonian.
It would be interesting to contrast the role of the
`induced correlations' in our cavity-emitter system
which arise from mixed photon and emitter
statistics, with the correlations arising from
electron-electron interactions in the
periodic Anderson Hamiltonian (which has
only fermionic particles).

\vskip0.20in \noindent
\underbar{Acknowledgements}
J.M.~and A.B.~were supported by the Research Experience for Undergraduates
program (NSF grant PHY-1852581).  T.C. is a McNair and MURPPS scholar
at the University of California, Davis.  R.T.S.~was supported by the grant DE‐SC0014671 funded by the U.S. Department of Energy, Office of Science. M.R. was supported by the National Science Foundation CAREER award 2047564.

\newpage

\bibliography{QST}

%%%%%%%%%%%%%%%%%%%%%%%%%%%%%%%%%%%
%%%%%%%%%%%%%%%%%%%%%%%%%%%%%%%%%%%
\newpage
\section{Appendix}
%%%%%%%%%%%%%%%%%%%%%%%%%%%%%%%%%%%
%%%%%%%%%%%%%%%%%%%%%%%%%%%%%%%%%%%

\vskip0.10in \noindent
{\it i.  Constraint on parity of $N$ to solve the IEP}

\vskip0.05in 
Our Monte Carlo solution to the IEP to determine 
$\{ \, J_i \, \}$  and
$\{ \, g_i \, \}$ 
for $N$ cavities each with one emitter
worked only for $N$ even.  This is because for odd $N$
the parities of the number of cavities, $N$, and 
the number of cavities+emitters, $2N$, are different.  
More precisely, when $N$ is odd there is a zero eigenvalue in the
spectrum of Eq.~\ref{eq:Jengineered}.  We cannot reproduce this zero
with our procedure of using the the cavity-only $2N$ spectrum
as the target for the $N+N$ cavity-emitter spectrum.

\begin{figure}[t] 
\includegraphics[width=1\columnwidth]{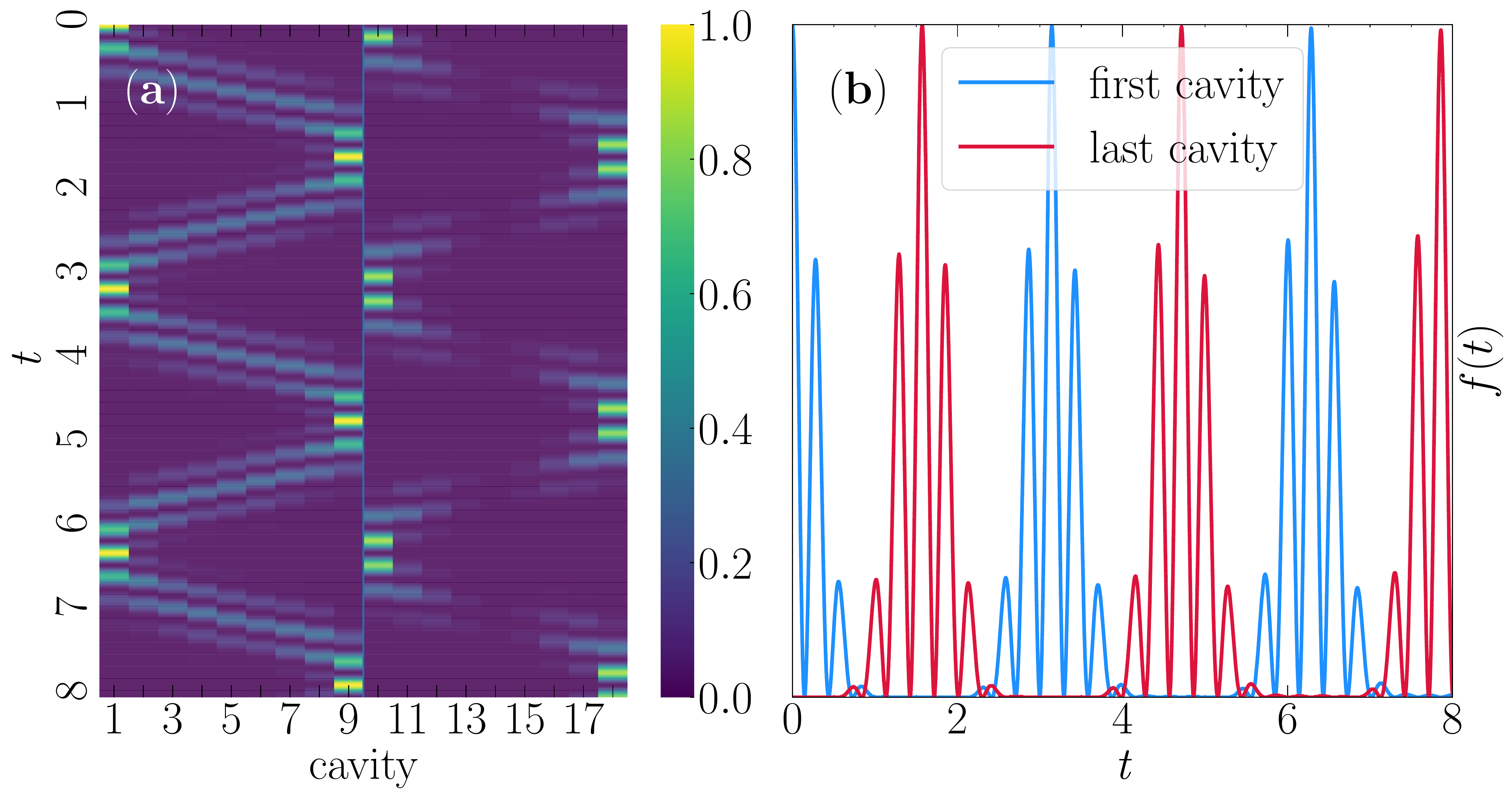}
\caption{We consider a sparse JCHH system of nine cavities and eight emitters, in which all cavities except the center cavity contain an emitter. 
Target eigenvalues were chosen to be those giving perfect QST for
a $N=17$ cavity-only chain.
In panel
 \textbf{(a)} we graph the probability that the photon is in each cavity and
 in each emitter for multiple times. The eighteen columns on the x axis
 represent the nine cavities and eight emitters, with the center emitter column left as zero. Time descends
 from 0 to 8 along the y axis. For every time and location, the
 probability is indicated in the color
 bar. In panel \textbf{(b)} we display the probability in starting and
 receiving cavities as a function of time. We observe perfect
 QST at time $\pi/2$ and with period $\pi$.}
\label{fig:QST98}
\end{figure}

To test this constraint on solvability further, we attempt a Monte Carlo 
solution for odd $N$, but removing
the emitter in the central cavity, so that the number of cavities+emitters, $2N-1$
is now also odd.
Results are given in Fig.~\ref{fig:QST98} and demonstrate that (near) perfect QST is
recovered.

\begin{center}
\begin{table}[t!]
\renewcommand{\arraystretch}{1.2}
\begin{tabular}{|c|c|c|c|c|}
\hline
 & \multicolumn{2}{|c|}{$N=N_e=12$} & \multicolumn{2}{|c|}{$N=N_e=16$} \\
\hline
Bond $i$ &  $J_i$ & $g_i$ & $J_i$ & $g_i$  \\
\hline
 1  & 5.597 & 14.755 & 6.519 & 19.929 \\
\hline
 2  & 7.712 & 13.056 & 9.030 & 18.264 \\
\hline
 3  & 9.255 & 11.278 & 10.876 & 16.511 \\
\hline
 4  & 10.322 & 9.261 & 12.310 & 14.708 \\
\hline
 5  & 11.355 & 7.025 & 13.515 & 12.753 \\
\hline
 6  & 12.007 & 3.987 & 14.461 & 10.581 \\
\hline
 7  & 11.355 & 3.987 & 15.294 & 8.0311 \\
\hline
 8  & 10.322 & 7.025 & 16.003 & 4.6159 \\
\hline
 9  & 9.255 & 9.261 & 15.294 & 4.6159 \\
\hline
 10 & 7.712 & 11.278 & 14.461 & 8.0311 \\
\hline
 11 & 5.597 & 13.056 & 13.515 & 10.581 \\
\hline
 12 &   & 14.755 & 12.310 & 12.753 \\
\hline
 13 &   &  & 10.876 & 14.708 \\
\hline
 14 &   &  & 9.0301 & 16.511 \\
\hline
 15 &   &  & 6.519 & 18.264 \\
\hline
 16 &   &  &  & 19.929 \\
\hline
\end{tabular}
\caption{
Intercavity hoppings $J_i$ and cavity-emitter couplings
$g_i$ which give perfect QST for $N=12$ and $N=16$ length
cavity arrays with a single emitter in each cavity.
}
\label{table:moreJg}
\end{table}
\end{center}

\begin{center}
\vskip0.10in
\begin{table}[ht!]
\renewcommand{\arraystretch}{1.2}
\hskip0.95in
\begin{tabular}{|c|c|c|c|c|}
\hline
Bond $i$ &  $J_i\,$: MC & $J_i$: Eq.~\ref{eq:empfitJ} 
 &  $g_i\,$: MC & $g_i$: Eq.~\ref{eq:empfitg}  \\
\hline
 1   & 4.521 & 4.527 & 9.558 & 9.562 \\
\hline
 2   & 6.158 & 6.164 & 7.824 & 7.806 \\
\hline
 3   & 7.232 & 7.246 & 5.872 & 5.826 \\
\hline
 4   & 7.979  & 8.000 & 3.234 & 3.231 \\
\hline
 5   & 7.232 & 7.246 & 3.234 & 3.231 \\
\hline
 6   & 6.158 & 6.164 & 5.872 & 5.826 \\
\hline
 7   & 4.521 & 4.527 & 7.824 & 7.806 \\
\hline
 8   &   &   & 9.558 & 9.562 \\
\hline
\end{tabular}
\caption{Comparison of the Monte Carlo (MC) and empirical cavity-cavity hoppings. We can see that the empirical formula,
Eq.~\ref{eq:empfitJ}, 
for $J_i$ is extremely accurate to the Monte Carlo derived $J_i$, but the formula for $g_i$,
Eq.~\ref{eq:empfitg},
is less accurate. This may be because the actual form for $g_i$ is more complex than our current fitting form.}
\label{table:empfit}
\end{table}
\end{center}

\vskip0.10in \noindent
{\it ii.  Additional data for perfect QST in the JCHH}

\vskip0.05in
Since a primary result of this paper is the computation
of $\{\, J_i, g_i \, \}$ which result in perfect QST 
for cavity-emitter systems, described by the JCHH, we provide in 
Table \ref{table:moreJg} some additional
results for large size systems, $N=12, 16$ to complement
the $N=8$ data provided in the main text.

\vskip0.50in \noindent
{\it iii.  Functional form for perfect QST JCHH couplings}

\vskip0.05in 
In the case of the cavity-only (spin chain), precise formulae for the
intercavity-hopping (Heisenberg exchange) constants to achieve perfect
QST are known.  The earliest
example is that of Christandl and given by  Eq.~\ref{eq:Jengineered}.
In the main manuscript we described a Monte Carlo process which works
in the more general cavity-emitter geometry.  However, this solution
is a `black box' in the sense that it produces {\it raw numbers} which achieve
(near) perfect QST without providing analytic insight or a formula.

We have attempted to fit the raw data produced by the simulation
to simple functional forms.  We mimic the spin-chain solution 
 \cite{christandl2004}
 with an {\it ansatz} of the square root of a polynomial function
 on $N$ and $i$. Indeed, the data collected allows a good fit to 
 the empirical formulae: 
\begin{equation}
    J_i=\frac{\sqrt{i(11N-6i)}}{2}
    \label{eq:empfitJ}
\end{equation}
\begin{equation}
    g_i=\frac{\sqrt{(2i-N-1)(14i-27N-7)}}{4}
    \label{eq:empfitg}
\end{equation}

Table \ref{table:empfit} compares the Monte Carlo values
with these empirical formulae.

\vskip0.10in \noindent
%% \underbar{iv.  Criterion for perfect QST}
{\it iv.  Criterion for perfect QST}

\vskip0.05in 
We note that it is non-trivial to distinguish whether small deviations
from fidelity ${\cal F} \equiv 1$ arise from a fundamental inability
to achieve perfect QST or from small randomness in the Monte
Carlo evaluation of the couplings.  We use the term `perfect QST' when 
our numerics indicate that by systematically running longer we can 
achieve arbitrarily close to ${\cal F} \equiv 1$.  In principle an extrapolation
of ${\cal F}$ as a function of simulation time would provide a more rigorous
analysis.  We do not do this here, because such an extrapolation
is complicated by the necessity to tune the annealing protocol, i.e.~the
manner in which $\beta$ is increased, as well as the choices for
the initial $\beta_i$ and final $\beta_f$.  We therefore elect to use a more
loose definition of `perfect QST', ${\cal F}$ very close to 1 and
systematically improvable.

\end{document}